\documentclass[a4paper]{article}

\usepackage{amsfonts}

\def\be{\begin{equation}}
\def\ee{\end{equation}}
\def\bea{\begin{eqnarray}}
\def\eea{\end{eqnarray}}
\def\({\left(}
\def\){\right)}
\def\<{\left<}
\def\>{\right>}

\def\[{\left[}
\def\]{\right]}

\def\be{\begin{equation}}
\def\ee{\end{equation}}
\def\bea{\begin{eqnarray}}
\def\eea{\end{eqnarray}}
\def\({\left(}
\def\){\right)}
\def\<{\left<}
\def\>{\right>}

\def\[{\left[}
\def\]{\right]}

\def\+{\bar}
\def\mb{\mathbb}

\def\Tr{{\mbox{Tr}}}

\def\t{\tilde}
\def\A{{\cal{A}}}
\def\P{{\cal{P}}}
\def\q{{\cal{O}}}

\begin{document}

\pagestyle{empty}
\vskip-10pt
\vskip-10pt
\hfill 
\begin{center}
\vskip 3truecm
{\Large \bf
Selfdual strings and loop space Nahm equations}\\ 
\vskip 2truecm
{\large \bf
Andreas Gustavsson}\footnote{a.r.gustavsson@swipnet.se}\\
\vskip 1truecm
{\it F\"{o}rstamajgatan 24,\\
S-415 10 
G\"{o}teborg, Sweden}\\
\end{center}
\vskip 2truecm
{\abstract{We give two independent arguments why the classical membrane fields should be take values in a loop algebra. The first argument comes from how we may construct selfdual strings in the M5 brane from a loop space version of the Nahm equations. The second argument is that there appears to be no infinite set of finite-dimensional Lie algebras (such as $su(N)$ for any $N$) that satisfies the algebraic structure of the membrane theory.}}
\vfill 
\vskip4pt
\eject
\pagestyle{plain}

\section{Introduction}
This paper concerns the selfdual string soliton in the M5 brane. We assume a straight string with $SO(4)$ rotational symmetry, and we wish to analyze its shape for $SU(2)$ gauge group proceeding by analogy with the magnetic monopole construction in super Yang-Mills theory. 

Having strings with $SO(4)$ rotational symmetry, it is very natural to expect that the equation that corresponds to the Nahm equation in the construction of magnetic monopoles, should also have this $SO(4)$ symmetry. Moreover, it is natural to suspect that a fuzzy three-sphere would play a similar role for the selfdual string construction as the fuzzy two-sphere plays for the magnetic monopole. 

More precisely, for monopoles one has solutions to the Nahm equation of the form
\bea
T^I \sim \frac{t^I}{z-1}
\eea
near $z=1$, where $t^I$ are coordinates on a fuzzy two-sphere. That is, $t^I$ are some representation matrices for $SU(2)$. For the selfdual string the analog of this would be
\bea
T^i \sim \frac{G^i}{\sqrt{z-1}}
\eea
near $z=1$, where $G^i$ now are coordinates on a fuzzy three-sphere. This approach was taken in \cite{Basu-Harvey, Berman:2006eu} among others.

The construction of the fuzzy three-sphere \cite{Ramgoolam:2001zx} is complicated compared to the fuzzy two-sphere. But if one notices the isomorphism $su(2) \oplus su(2) \simeq so(4)$, one could think that the fuzzy three sphere could be described in terms of (or at least be mapped to) two fuzzy two-spheres. 

If $G^i$ denote the coordinates on a fuzzy three-sphere in a certain reducible representation of $SO(4)$, characterized by the single integer $n$ (see \cite{Ramgoolam:2001zx} for details), and if we define $G^{ij} = \frac{1}{2}[G^i,G^j]$, and let $\lambda^I_{ij}$ denote the 't Hooft matrices (defined in Appendix \ref{A}), then 
\bea
t^I &=& \frac{1}{2}\lambda^I_{ij} G^{ij}\cr
\t t^I &=& \frac{1}{2}\t \lambda^I_{ij} G^{ij} 
\eea
can be computed, for instance from Eq (17) in \cite{Berman:2006eu}.\footnote{Eq (17) in \cite{Berman:2006eu} reads
\bea
G^{ij}\P_{\pm} &=& \sum_{r=1}^n\(\frac{n+3}{4}\rho_r(\gamma^{ij}P_{\pm}) - \frac{n+1}{4} \rho_r(\gamma^{ij}P_{\mp})\)\P_{\pm}
\eea
(For $n=1$, this is $G^{ij} = \gamma^{ij}$). From this we get
\bea
\lambda^I_{ij} G^{ij} \P_+ &=& (n+3)\sum_r\rho_r(\sigma^I)\cr
\lambda^I_{ij} G^{ij} \P_- &=& -(n+1)\sum_r\rho_r(\sigma^I).
\eea
Here $4\sigma^I = \lambda^I_{ij} \gamma^{ij}$. Then the sum $\lambda^I_{ij} G^{ij} \equiv 2t^I$ becomes
\bea
t^I &=& \sum_r \rho_r(\sigma^I).
\eea} The result is very simple,
\bea
t^I &=& \sum_r \rho_r(\sigma^I)\cr
\t t^I &=& \sum_r \rho_r(\t \sigma^I)
\eea
where $[\sigma^I,\sigma^J] = 2i\epsilon_{IJK}\sigma^K$ are the Pauli sigma matrices. There is a tensor product and $\rho_r(\sigma^I)$ means $1\otimes \cdots\otimes \sigma^I\otimes \cdots \otimes 1$ where $\sigma^I$ is placed at position $r=1,...,n$. This shows that there is a map from the fuzzy three-sphere coordinates to the coordinates $t^I$ and $\t t^I$ of two fuzzy two-spheres,
\bea
4it^I &=& \epsilon_{IJK} [t^J, t^K]\cr
4i\t t^I &=& \epsilon_{IJK} [\t t^J, \t t^K]
\eea
This suggests that it might be possible to describe the fuzzy three-sphere in this alternative way as two fuzzy two-spheres. 

One could now also suspect that the selfdual string can be viewed in some sense as two decoupled sets of magnetic monopoles. This is what we wish to make precise in this paper. 

We will work in loop space, and use loop space fields, that is, fields that depend not on a point but on a whole loop $C$ embedded in space. In particular we have a loop space gauge field $A_{\mu s}(C)$ which, in the abelian case, may be given the ultra local form
\bea
A_{\mu s}(C) &=& B_{\mu\nu}(C(s)) \dot{C}^{\nu}(s)\label{ultra}
\eea
where $B$ denotes the usual two-form gauge field, which is one of the fields in the tensor multiplet in six dimensions.

In \cite{Gustavsson:2005aq} it was shown that $(2,0)$ supersymmetry can be extended to non-abelian loop space fields under quite general assumptions. One need not assume ultra locality for this to work. Neither do we need any assumptions on ultra locality in what we do in this paper. We wish to leave ultra locality as an open crucial question. Eventually, if one wants to make concrete computations, one will have to face the question. For the time being we content ourselves with setting up some algebraic framework in loop space, that should put some constraints on what $(2,0)$ theory could be. Loop space is a very big space, and it seems to us like $(2,0)$ theory could live somewhere in it. But we do not know exactly where it lives. More precisely, one should find the appropriate constraints on the loop space fields.

In the abelian case, we notice that we need the three-form field strength $H_{\mu\nu\rho}$ to be anti-self-dual in order to close two supersymmetry variations into a superalgebra on the equations of motion. However, as was seen in \cite{Gustavsson:2005aq}, we do not need to enforce self-duality on the corresponding loop space field strength $F_{\mu s,\nu t}$ in order to close supersymmetry on-shell. Intuitively one can understand why this should be so, by inserting the condition of self-duality into the loop space field strength. We then find that
\bea
F_{\mu s,\nu t}(C) &=& H_{\mu\nu\rho}(C(s))\dot{C}^{\rho}(s)\delta(s-t)\cr
&=& \frac{1}{6}\epsilon^{\kappa\tau\sigma}{}_{\mu\nu\rho}H_{\kappa\tau\sigma}(C(s))\dot{C}^{\rho}(s)\delta(s-t)
\eea
but the second line in this equation can not be expressed in terms of $F_{\mu s,\nu t}(C)$, which shows that we can not implement self-duality on $F_{\mu s,\nu t}(C)$.\footnote{Possibly one may relate $F(C)$ to some sort of hodgedual of $F(C')$ when evaluated at a different loop $C'$ so that one would have a non-local self-duality condition in loop space. Now supersymmetry acts locally in loop space and relates the variation of a bosonic (fermionic) field at $C$ to fermionic (bosonic) fields evaluated at the same point $C$, hence any non-local constraints on loop space fields do not concern us when we study supersymmetry.} Happily, in \cite{Gustavsson:2005aq} we could show that no `self-duality' on $F_{\mu s,\nu t}$ is needed in order to close supersymmetry on-shell. Self-duality comes out from the loop space supersymmetry formalism once we insert the ultra local equation Eq (\ref{ultra}) for the loop space fields into the loop space supersymmetry variations. 

In \cite{Girelli:2003ev}, \cite{Baez:2005qu} a gerbe formalism was developed with ultra local expressions for non-abelian loop space gauge fields as defined in \cite{Schreiber:2005ff} based on ideas in \cite{Hofman:2002ey}. This formalism uses a local connection two-form $B_{\mu\nu}$ as well as a local connection one-form $A_{\mu}$ subject to vanishing fake curvature\footnote{Something like $F_{\mu\nu} \sim B_{\mu\nu}$ where $F$ is the curvature of $A$, though generically $F$ and $A$ can take values in different internal algebras and one must map $B$ to the same algebra as $A$ to make sense of this equation.}. Vanishing fake curvature was shown in \cite{Girelli:2003ev}, \cite{Baez:2005qu} is needed in order for Wilson surfaces to be well-defined. Presumably there are corresponding ultra local expressions for the other loop space fields in the tensor multiplet as well. It may be that one just need to hang on a gauge index on the scalars and fermions. Inspired by how things work in the abelian case, one could try and play the same game with such non-abelian ultra local loop space fields. Plugging such loop space fields into the supersymmetry variations in loop space obtained in \cite{Gustavsson:2005aq}, one can derive corresponding equations for the local fields (for the $B_{\mu\nu}$ and $A_{\mu}$). From such a computation one should find a non-abelian counterpart of the self-duality condition on $B_{\mu\nu}$. But now, just like self-duality could not be imposed in loop space, vanishing fake curvature can probably not either. But again, just like self-duality comes out from supersymmetry when we assume ultra locality in the abelian case, in the non-abelian case the condition of vanishing fake curvature could also come out from supersymmetry. Though this is just a speculative idea.

It would of course be extremely nice if an ultra local approach could be made to work consistently with supersymmetry, since then one could derive equations of motion for local space-time fields from loop space equations, and possibly also start to do real physics computations in non-abelian $(2,0)$ theory. 

Either way, we think that any possible set of equations that contains $(2,0)$ theory one way or the other, should be worth to study, even if it may be difficult to tell at the moment what these equations could be used for. Eventually $(2,0)$ theory (if eventually properly understood) could be useful in describing non-perturbative effects in QCD \cite{Witten:1997ep}, as well as giving a better understanding of M-theory and quantum gravity via AdS-CFT correspondence.

The redaction is as follows: In section \ref{one} we introduce suitable coordinates in loop space, that generalizes the Hopf map $S^3 \mapsto S^2$. In section \ref{two} we derive the Bogomolnyi equation for selfdual strings in terms of these new loop space coordinates. In section \ref{three} we present an abelian solution and a loop space generalization of the ADHMN construction of selfdual strings. In section \ref{four} we show how the corresponding Nahm equations can be obtained from the membrane theory. In section \ref{five} we show that this membrane theory can be reduced to super Yang-Mills theory. In section \ref{six} we show the relation between two Nahm equations and the Basu-Harvey equation in Ref \cite{Basu-Harvey}.

\section{Various coordinates in loop space}\label{one}
We assume static straight strings in $1+5$ dimensions. Hence we have translational invariance in $1+1$ dimensions, and we will restrict our attention to the transverse space to the strings, ${\mb{R}}^4$, with cartesian coordinates $x^i$. We then also consider the loop space over ${\mb{R}}^4$. If we parametrize the embedding of a loop in ${\mb{R}}^4$ as $s\mapsto C^i(s)$, we have coordinates for this loop that are $C^i(s)$ with $s$ being a continuous index. Hence loop space is an infinite-dimensional space.

If the loop is planar, then we can compute the area that it encloses by integrating 
\bea
\sigma^{ij}_s := C^i(s) \dot{C}^j(s) - C^j(s) \dot{C}^i(s)
\eea
around the loop. But we now define this quantity for any loop, and then we may define a new set of coordinates on loop space as\footnote{For the zero mode part this is the Hopf map from $S^3$ to $S^2$.}
\bea
X^I_s &:=& \frac{1}{2}\lambda^I_{ij}\sigma_s^{ij}\cr
\t X^I_s &:=& \frac{1}{2}\t \lambda^I_{ij}\sigma^{ij}_s
\eea
that we may invert to get
\bea
\sigma^{ij}_s &=& -\frac{1}{2}\(\lambda^I_{ij} X^I_s + \t \lambda^I_{ij}\t X^I_s\)
\eea
We now claim that either $X$ or $\t X$ can be used as coordinates on loop space. To see this we would like to invert the maps
\bea
C^i(s) &\mapsto &  X^I_s\cr
C^i(s) &\mapsto & \t X^I_s
\eea
but this seems to be very difficult, by making an exact computation. So instead we consider a wavy line in Monge gauge
\bea
C^i(s) = (s, \xi^a(s))
\eea
where $a=1,2,3$, and compute the inverse just to lowest order in the fluctuations about a straight line. We get
\bea
\sigma^{1a}_s &=& s^2 \frac{d}{ds} (s^{-1}\xi^a(s))
\eea
and from this, we can obtain $\xi^a(s)$ in terms of $X^I_s$ by means of an integration. Then we get 
\bea
\frac{\partial C^i(s)}{\partial X^I_t} &=& -\frac{s}{t^2}\theta(s-t) \lambda^I_{1i} + \q(\xi)\cr
\frac{\partial C^i(s)}{\partial \t X^{I}_t} &=& -\frac{s}{t^2}\theta(s-t) \t \lambda^{I}_{1i} + \q(\xi).
\eea
Here $\theta(\bullet)$ denotes the Heaviside step function. We can now check that these constitute the inverse mappings to 
\bea
\frac{\partial X^I_t}{\partial C^i(u)} &=& \lambda^I_{1i}t^2 \frac{d}{dt}\(t^{-1}\delta(t-u)\) + \q(\xi)\cr
\frac{\partial X^I_t}{\partial \t C^i(u)} &=& \t \lambda^I_{1i}t^2  \frac{d}{dt}\(t^{-1}\delta(t-u)\) + \q(\xi)\label{inv}
\eea
in the sense that
\bea
\int dt \frac{\partial C^i(s)}{\partial X^I_t} \frac{\partial X^I_t}{\partial C^j(u)} &=&  \delta_{ij}\delta(s-u)\cr
\int dt \frac{\partial C^i(s)}{\partial \t X^I_t} \frac{\partial \t X^I_t}{\partial C^j(u)} 
&=& \delta_{ij}\delta(s-u),\cr
\int ds \frac{\partial C^i(s)}{\partial X^I_t} \frac{\partial X^J_r}{\partial C^i(s)} &=& \delta_{IJ}\delta(t-r)
\eea
We may also note that 
\bea
\frac{\partial X^I_s}{\partial \t X^J_t} = \int dr \frac{\partial X^I_s}{\partial C^i(r)} \frac{\partial C^i(r)}{\partial \t X^J_t} \sim (\lambda^I\t \lambda^J)_{11} \eea
is a non-singular matrix. Hence $X\mapsto \t X$ is a well-behaved coordinate transformation.

We have seen that it is always possible to get $C^i(s)$ from $\sigma^{ij}_s$ (in fact from $\sigma^{i1}_s$ in Monge gauge) by means of an integration. We have also seen that $X^I_s$ and $\tilde X^I_s$ are dependent coordinates. We may probably always choose either $X^I_s$ or $\t X^I_s$ as independent coordinates, in place of (gauge fixed) coordinates $C^i(s)$. If we impose a gauge fixing on the parametrization of $C^i(s)$, then we have three independent coordinates that describes the loop. But such a gauge fixing on $C^i(s)$ does not imply any gauge fixing on $X^I_s$. So now we do have a matching number of coordinates.

\section{The Bogomolnyi equation}\label{two}
In the abelian case, the Bogomolnyi equation for selfdual strings was obtained in \cite{Howe:1997ue}. It is given by
\bea
H_{ijk} = \epsilon_{ijlk}\partial_l\phi
\eea
where $H_{ijk} = \partial_i B_{jk} + \partial_j B_{ki} + \partial_k B_{ij}$ is the gauge field strenght of the two-form gauge potential $B_{ij}$.

To generalize this to the non-abelian case, we first introduce abelian loop space fields,
\bea
A_{is} &=& B_{ij}(C(s))\dot{C}^j(s)\cr
\phi_{is} &=& \dot{C}_i(s)\phi(C(s))
\eea
and define the field strength $F_{is,jt} = \frac{\delta}{\delta C^i(s)}A_{jt}-\frac{\delta}{\delta C^j(t)}A_{is}$. We get
\bea
F_{is,jt} &=& H_{ijk}(C(s))\dot{C}^j(s)\delta(s-t).\label{H}
\eea
We then find that the Bogomolnyi equation can be written as
\bea
F_{is,jt} = \epsilon_{ijkl} \partial_{ks}\phi_{lt}
\eea
where $\partial_{is} := \frac{\delta}{\delta C^i(s)}$.

We now propose that the non-abelian generalization of this equation is given by \cite{Gustavsson:2006ie}
\bea
F_{is,jt} = \epsilon_{ijkl} D_{ks}\phi_{lt}.\label{self}
\eea
where $D_{is} = \partial_{is} + A_{is}$ is the gauge covariant derivative.

We can write the one-form gauge potential in loop space 
\bea
\A(C) = \int ds A_{is} \delta C^i(s)
\eea
in the various coordinate systems as follows,
\bea
A(C) &=& \int ds A_{I}(s,X) \delta X^I_s\cr
&=& \int ds \tilde{A}_{I}(s,\t X) \delta \t X^I_s
\eea
From this, we get that
\bea
A_{is} = \lambda^I_{ij}\(A_{I}(s)\dot{C}^j(s)+\frac{1}{2}\dot{A}_{I}(s)C^j(s)\)\label{wrong}
\eea
In concordance with this, we also let
\bea
\phi_{is}(C) = \dot{C}_i(s)\phi(s,X) + \frac{1}{2}C_i(s)\dot{\phi}(s,X)
\eea
for the scalar field.

In \cite{Gustavsson:2006ie} it was shown that the $SU(2)$ covariant Bogomolnyi equation
\bea
\frac{1}{2}\epsilon_{IJK}F_{IJ}(s,t) &=& D_K(s) \phi(t),\label{first}
\eea
with the above definitions of the fields, implies the $SO(4)$ covariant Bogomolnyi equation, Eq (\ref{self}), for selfdual strings. Here $D_K(s) = \frac{\delta}{\delta X^K_s} + A_K(s)$. This however, does not show that any solution to Eq (\ref{self}) can be obtained solely from Eq (\ref{first}), and in fact this is not true. We need another copy of the $SU(2)$ equation.

We can choose to express a field in terms of the coordinates $X$ or the coordinates $\t X$. Let now $A=A(X)$ and $\t A= \t A(\t X)$ commute, $[A,\t A] = 0$, and similarly $\phi=\phi(X)$ and $\t \phi = \t \phi(\t X)$ also commute, $[\phi,\t \phi]=0$. Then we may consider the two Bogomolnyi equations for these fields,
\bea
F_{IJ} &=& \epsilon_{IJK} D_K \phi\cr
\t F_{IJ} &=& \epsilon_{IJK} \t D_{K} \t \phi
\eea
and find that  
\bea
A_{is} &=& \partial_{is} X^I_t A_I + \partial_{is} \t X^{\t I} \t A_{\t I} \cr
\phi_{is} &=& \dot{C}_i(s) \(\phi + \t \phi\) + \frac{1}{2} C_i(s) \(\dot{\phi} + \dot{\t \phi}\)
\eea
satisfies the Bogomolnyi equation. 

Connection with local physics is provided by letting
\bea
B_{ij}(s) = \lambda^I_{ij} A_I(s) + \lambda^{\t I}_{ij} A_{\t I}(s)
\eea
for the two-form gauge potential. This relation can now be inverted, to express $A_I$ and $\t A_I$ in terms of $B_{ij}$ or in terms of $A_{is}$. 

In the abelian case, we can then let $B_{ij}(s,C) = B_{ij}(C(s))$ be the usual local two-form gauge potential. For point-like loops our definition coincides with the conventional definition
\bea
A_{is} = B_{ij}(C(s))\dot{C}^j(s).
\eea

We now show the the $SO(4)$ Bogomolnyi equation implies these two $SU(2)$ Bogomolnyi equations. We compute 
\bea
\epsilon_{IJK}F_{IJ}(s,t) &=& \epsilon_{IJK}\int du\int dv \frac{uv}{s^2t^2}\lambda^I_{1i}\lambda^J_{1j}\theta(u-s)\theta(v-t)F_{iu,jv}\cr
&=& \frac{1}{s^2t^2}\epsilon_{IJK}\lambda^I_{1i}\lambda^J_{1j}\epsilon_{ijkl}\int du\int dv uv \theta(u-s)\theta(v-t)D_{ku}\phi_{lv}\cr
&=& 2\lambda^K_{1i} \frac{1}{s^2t^2} \int du\int dv uv\theta(u-s)\theta(v-t)D_{iu}\phi_{1v}
\eea
We note that 
\bea
\phi_{1s} &=& \phi(s) + \frac{1}{2}s\dot{\phi}(s)
\eea
in Monge gauge. Then we use
\bea
D_{iu} &=& \int dw \frac{\partial X^I_w}{\partial C^i(u)} D_I(w)
\eea
and get
\bea
\epsilon_{IJK}F_{IJ}(s,t) &=& 2\lambda^K_{1i}\lambda^I_{i1}\frac{1}{s^2t^2} \int du\int dv \int dw uv\cr
&&\theta(u-s)\theta(v-t)\(\delta(w-u) + w\dot{\delta}(u-w)\) \cr&&D_I(w)\(\phi(v)+\frac{1}{2}v\dot{\phi}(v)\)\cr
&=& 2D_K(s)\phi(t)
\eea
The same type of computation can be done to show that
\bea
\epsilon_{IJK}\t F_{IJ}(s,t) &=& 2\t D_K(s)\t \phi(t).
\eea

\subsection{Constructing selfdual string solutions}\label{three}
In the abelian case, far away from a selfdual string, the $U(1)$ field strength is of the form 
\bea
H_{ijk}(x) = \epsilon_{ijkl} \frac{x^l}{|x|^4}.\label{string}
\eea
The string being self-dual means that $H_{i05} \sim \epsilon_{ijkl}H_{jkl}$. From now on we will only focus on the components $H_{ijk}$ of the field strength, which means that we may forget about self-duality. From Eqs (\ref{string}), (\ref{H}) we then get
\bea
F_{is,jt} = \epsilon_{ijkl}\frac{\sigma^{kl}_s}{(R_s)^4} \delta(s-t)
\eea
where $R_s=|C^i(s)|$. We would now like to transform this solution into our new coordinates in loop space. However, we do not get any nice (i.e. symmetric) expression if we just use the $X$ coordinates. We can get a much nicer expression if we use both $X$ and $\t X$ coordinates. We then first separate $F_{is,jt}$ into selfdual and antiself dual pieces\footnote{By selfdual we simply mean with respect to $\epsilon_{ijkl}$, and not with respect to some strange loop space hodgeduality operator. It should be noted that this decomposition into self-dual pieces has nothing to do with self-duality of $H_{\mu\nu\rho}$!} as
\bea
F_{is,jt} = F_{is,jt}^+ + F_{is,jt}^-.
\eea
We then apply the wavy line approximation. Now to be slightly more general, and to get $s$ dimensionless, we introduce parameters $R^a$ of dimension length, and parametrize the wavy line as
\bea
C^i(s) = (Rs, R\xi^a(s) + R^a)
\eea
where $R=\sqrt{R^aR_a}$ may be thought of as a radius. Then we find that
\bea
X^I_s X^I_s = R^4 (1+ \q(s,\xi))
\eea
Then noting the identity $(\lambda^I\lambda^J\lambda^K)_{11} = i\epsilon_{IJK}$, we get
\bea
F_{IJ}(s,t,X) &=& \frac{1}{R^2}\epsilon_{IJK} \int du \frac{u^2}{s^2t^2} \theta(u-s)\theta(u-t)\frac{X^K_u}{(R_u)^4}\label{f1}
\eea
from the $F^+$ piece, and similarly, 
\bea
\t F_{IJ}(s,t,\t X) &=& \frac{1}{R^2}\epsilon_{IJK} \int du \frac{u^2}{s^2t^2} \theta(u-s)\theta(u-t)\frac{\t X^K_u}{(R_u)^4}\label{f1}
\eea
from $F^-$. These solutions look rather strange, but if we write $s^{-2} = -\partial_s(s^{-1})$, and make `an integration by parts' by moving the derivative $\partial_s$ to the theta function (eventhough there is no integration over $s$), and then do the same thing for $t^{-2}$, then we get
\bea
F_{IJ}(s,t,X) &=& \frac{1}{R^2}\epsilon_{IJK} \frac{X^K_s}{(R_s)^4}\delta(s-t)\cr
&=& \epsilon_{IJK} \frac{X^K_s}{|X_s|^3}\delta(s-t) (1+ \q(\xi))
\eea
which equals the field strength from a Dirac monopole (in a way that is consistent with the accuracy of our approximation)\footnote{Now this is up to a total derivative. But it seems likely that this total derivative term could appear in the $\q(\xi)$ as well.}.

If we would not use both $X$ and $\t X$ coordinates, we would not get a solution that would resemble a Dirac monopole at all. This is why we could not get any selfdual string solution by solving just one $SO(3)$ covariant Bogomolnyi equation -- we need {\sl{two}} copies of this equation. In general we need to solve these two decoupled copies of this equation and then look for the physical string solution as a certain linear combination of these solutions.

The simplest non-abelian solution to the $SO(3)$ Bogomolnyi equation is
\bea
\phi(s,X) &=& \varphi(X_s)\cr
A_{\mu}(s,X) &=& {\cal{A}}_{\mu}(X_s)
\eea
where $\varphi(x) \sim 2\coth (2x) - \frac{1}{x}$ is the celebrated Higgs profile of the 't Hooft-Polyakov monopole, and ${{\cal{A}}}_{\mu}(x)$ may be taken as the hedgehog solution for the gauge field. 

This solution can be obtained by imitating the ADHMN construction. We then take the trivial solution $T^I(s) = 0$ to the Nahm equation (to be presented shortly). Then the construction equation (also to be presented shortly) reads
\bea
\(-\frac{\delta}{\delta z(t)} + X^I_t \sigma_I \) v &=& 0
\eea
where $z(s)\in [-1,1]$. This equation has solutions
\bea
v = N(X) P\exp \(\int dt z(t) X^I_t \sigma_I\) n
\eea
where $n$ is an element in an orthonormal basis of vectors. P denotes a path ordered exponent. We separate these solutions into factors as
\bea
v = \prod_s v_s\label{r1}
\eea
This decomposition becomes unique if we in addition impose the normalization conditions
\bea
\int_{-1}^1 dz(s) (v_s)^2 = 1\label{r2}
\eea
for each $s$. We then find that
\bea
v_s = N(s,X) \exp \(z(s) X^I_s \sigma_I\) n
\eea
The normalization conditions yield
\bea
N(s,X) = \sqrt{\frac{|X_s|}{\sinh(2|X_s|)}}
\eea
(Here $|\bullet|$ denotes the standard euclidean norm.) and then we get
\bea
\phi(s,X) &=& \prod_{r\neq s}\int dz(r) (v_r)^2 \int dz(s) z(s) (v_s)^2\cr
&=& N(X_s)^2\int_{-1}^1 dz(s) z(s) \exp \(2 z(s) X^I_s \sigma_I\).
\eea
The evaluation of this integral gives the 't Hooft-Polyakov solution presented above.

In an attempt to find the general solution to one of these $SO(3)$ Bogomolnyi equations, in the case of $SU(2)$ gauge group, we use the ADHMN construction for magnetic monopoles. We thus make the following ansatz
\bea
A_I(s) &=& -i\int [dz] \prod_t v^{\dag}_t \partial_I(s) v_t\cr
\phi(s) &=& \int [dz] \prod_t z(s) v_t^{\dag} v_t 
\eea
and we make a corresponding ansatz for the tilde fields $\t A$ and $\t \phi$. Here $v_t = v_t(z,X)$ and $z=z(s)$ is a one-dimensional loop. In the path integral we integrate over the range $z(s) \in [-1,1]$ for each $s$. We normalize $v_r$ as
\bea
\int_{-1}^1 dz(r) v_r^{\dag} v_r &=& 1
\eea
for each $r$, and we let $v_r$ be subject to the construction equations
\bea
\Delta^{\dag}_s v &=& 0,\label{norm}
\eea
where
\bea
\Delta_r(z) = \frac{\delta}{\delta z(r)} + \(X^I_r - T^I(z,r)\)\sigma^I
\eea
and where the $T^I$ obey the generalized Nahm equation
\bea
\frac{\delta T^K(s)}{\delta z(r)} + \frac{i}{2}\epsilon_{IJK}[T^I(s), T^J(r)]  &=& 0.
\eea
We then define $v_r$ by Eqs (\ref{r1}), (\ref{r2}).

To see that this really gives solutions to our Bogomolnyi equation, one can just make very small modification of existing derivations of the ADHMN construction. We present such a slightly modified derivation in the Appendix \ref{B}. 

Also one should specify the boundary conditions. By suitable choice of units (i.e. of the Higgs vev), we have already assumed that the range is $z(s) \in [-1,1]$ and we now give the boundary conditions as
\bea
T^I(s) = -\frac{t^I(s)}{z(s) \mp 1} + \q(1).
\eea
near $z(s) = \pm 1$. Then 
\bea
2i t^K(s)\delta(s-r) &=& \epsilon_{IJK} [t^I(s), t^J(r)]
\eea
From this we conclude that
\bea
[t^I(s),t^J(r)] &=& \epsilon_{IJK} \delta(s-r) t^K(s) + K\delta^{IJ}\dot{\delta}(s-r)
\eea
where we can allow for a central extension.

\section{Nahm's equations from the membrane}\label{four}
We now wish to see if we can obtain the Nahm equation as the Bogomolnyi equation for the $M2$ brane. If the $M5$ brane is extended in the directions $0,1,2,3,4,5$ and the M2 brane in the directions $0,1,6$ then the unbroken supersymmetries satisfy \cite{Bagger:2006sk}
\bea
\Gamma^6\epsilon=\Gamma^{2345}\epsilon
\eea
This brane configuration is translationally symmetric in all directions but the $6$-direction. In particular the branes intersect along a string aligned along the $1$-direction. We assume that only the scalar fields on M2 are excited which are parallel to M5 brane, and we denote these excited scalar fields as $X^i$ for $i=2,3,4,5$. The supersymmetry variation for the fermions on the $M2$-brane  (assuming the conjectured membrane theory in \cite{Bagger:2007jr}) is given by
\bea
\delta \psi = -\Gamma^6 \Gamma_i\epsilon D_6 X^i - \frac{1}{6} \Gamma_{ijk}\epsilon [X^i,X^j,X^k]_{M2}
\eea
The condition for unbroken supersymmetry can be rewritten as
\bea
\epsilon^{ijkl}\Gamma_l\Gamma^6 \epsilon = \Gamma^{ijk}\epsilon
\eea
By demanding $\delta \psi=0$ for such supersymmetry parameters, we find the Nahm equation
\bea
\partial_6 X^i \sim \epsilon_{ijkl}[X^j,X^k,X^l]_{M2}.
\eea
It seems like there is only one way one can extend the membrane theory if one wants to leave the $SO(4)$ gauge group. This is to let the gauge group be of the form $\hat{SU}(N)$, where $\hat{SU}(N)$ denotes the loop algebra of $SU(N)$ with a non-trivial central extension. More generally one could take the loop algebra extension of any semi-simple Lie algebra. We denote the generators as $T^a(s),1$, where $\Tr(T^a(s))=0$, and we choose a normalization such that $\Tr(1)=1$. We define the membrane three-bracket as \cite{Awata:1999dz}
\bea
[a,b,c]_{M2} = [a,b]\Tr(c) + [b,c]\Tr(a) + [c,a]\Tr(b).\label{bracket}
\eea
One may verify that this three-bracket satisfies the Fundamental Identity (the analog of the Jacobi identity for three-algebras, see \cite{Bagger:2007jr}). 

Let $t^I$ generate an $su(2)$ subalgebra of $\hat{su}(N)$. Using the membrane bracket, we then find that
\bea
[t^I,t^J,1]_{M2} &=& \epsilon^{IJK}t^K
\eea
and all the other brackets vanish, so for instance $[t^I,t^J,t^K]_{M2} = 0$.

If we take a gauge group that has $su(2)\oplus su(2)$ as a subalgebra, we can solve the Nahm equation by taking  
\bea
X^I(z) &=& T^I(z) + \t T^I(z)\cr
X^4(z) &=& 1
\eea
where $x^6$ is here denoted by $z$ as conventional, thereby descending to the Nahm equation 
\bea
\partial_6 T^I &=& \epsilon^{IJK} T^J T^K\cr
\partial_6 \t T^I &=& \epsilon^{IJK} \t T^J \t T^K.
\eea
This is the zero mode part of the Nahm equation that we found earlier in our construction of selfdual strings.

To get all modes from the membrane, we drop the integration over $s$, and let the  fields be (non-abelian) loops themselves, $X^I(s) = X^I_a(s)T^a(s)+X^I_{\sharp}(s)1$, $A_{\mu}(s) = A_{\mu,a}(s)T^a(s) + A_{\mu \sharp}(s)1$ (here we associate the index $\sharp$ to the central element). This in turn implies a covariant derivative $D_{\mu}(s)$ meaning that the fields must be functionals of loops $x^{\mu}(s)$. We assume a local dependence on these loops, as $X^I(s,[x])=X(x(s))$, $A_{\mu}(s,[x])=A_{\mu}(x(s))$. Then we take the supersymmetry variation as
\bea
\delta(t) X^I(s) &=& i\delta(s-t)\bar{\epsilon} \Gamma^I \psi(s)\cr
\delta(t) A_{\mu}(s) &=& \i \bar{\epsilon} \Gamma_{\mu} \Gamma_I [\psi(t),X^I(s),\bullet]\cr
\delta(t) \psi(s) &=& -\Gamma^{\mu}\Gamma_I \epsilon D_{\mu}(t) X^I(s) - \frac{1}{6}\Gamma_{IJK}\epsilon [X^I,X^J(t),X^K(s)].
\eea
The loop space supersymmetry algebra reads
\bea
[\delta_{\epsilon}(s),\delta_{\eta}(t)] &=& \delta(s-t) \bar{\epsilon}\Gamma^{\mu}\eta \partial_{\mu}(s)
\eea
and the supersymmetry variations close on the fermionic equation of motion
\bea
\Gamma^{\mu} D_{\mu}(s)\psi(t) + \frac{1}{2}\Gamma_{IJ}[\psi(s),X^I(t),X^J] &=& 0.
\eea

\subsection{Restriction to Yang-Mills}\label{five}
To see that this `membrane theory'\footnote{It is not known whether this theory really describes parallel M2 branes, though it is the only candidate.} contains Yang-Mills theory, we restrict to zero mode part and we let just one scalar field (let us choose it to be $X^{(8)}$, or in an eleven dimensional notation, the eights scalar field would be denoted as $X^{(10)}$) have a non-trivial central element $R$. We thus let
\bea
X^{(8)} = R + X^{(8)}_a T^a\label{R}
\eea
and we let the other scalars be of the form
\bea
X^M = X^M_a T^a
\eea
with no central element, for $I=M=1,...,7$. Then the three-bracket reduces to a commutator,
\bea
[X^M, X^N, X^8] = R [X^M,X^N].
\eea
The gauge field variation reduces to 
\bea
\delta A_{\mu} = \bar{\epsilon} \Gamma_{\mu}\Gamma_{(10)} \psi
\eea
Finally
\bea
\delta \psi = \Gamma^{\mu}\Gamma_I \epsilon D_{\mu}X^I + .. 
\eea
reduces to 
\bea
\delta \psi = \frac{1}{2}\Gamma^{\mu\nu}\Gamma_{(10)} \epsilon F_{\mu\nu} + \Gamma^{\mu}\Gamma_M \epsilon D_{\mu}X^M + ..
\eea
by dualizing the eight's scalar according to
\bea
D_{\mu}X^{(8)} = \frac{1}{2}\epsilon_{\mu\nu\rho} F^{\nu\rho}.\label{dual}
\eea
We thus take Eqs (\ref{dual}) and (\ref{R}) as a defining equation for the field $X^{(8)}$ that makes the membrane theory reduce to Yang-Mills theory. 

As motivation for this prescription we note that only one scalar field can be excited in the direction (that we labeled by coordinate $I=8$) of the compact M-theory circle, and that this compact dimension corresponds to a commuting coordinate, hence to an element in $U(1)$.

We note that supersymmetry enforces the M2 gauge field to contain no propagating degrees of freedom. When we reduce to D2, the gauge field somehow must acquire one degree of freedom. If we adopt the triple product in Eq (\ref{bracket}) for the M2 theory, it seems impossible to write a supersymmetric action because with this triple product we can not fulfil the invariance property $\Tr([x,y,a]b) + \Tr (a[x,y,b]) = 0$ of the trace form which is needed for a supersymmetric action \cite{Bagger:2007jr}. Still supersymmetry variations close on those equations of motion found in \cite{Bagger:2007jr}, but it seems there is no action from which they can be derived if we use this triple product.

This is a revised version of this paper, and here we would also like to see whether we could fit our restriction to Yang-Mills into the more general reduction procedure carried out in \cite{Mukhi}, \cite{Gran}. We define
\bea
X^I &=& \phi^I_a T^a + \varphi \cr
\psi &=& \psi_a T^a + \chi
\eea
where we have pulled out the $U(1)$ fields ($\varphi$ and $\chi$) explicitly. We call them $U(1)$ fields, though the actual Lie algebra structure of the theory is more intricate than just that of $SU(N)\times U(1)$. We discuss this more in Appendix $C$. It is clear the $U(1)$ fields obey free equations of motion because all interactions involve the three-bracket, and we can not get a central element from the three-bracket unless we have a non-trivial central extension. One could naively maybe think the gauge covariant derivative would not involve the three-bracket, but it does,\footnote{In the literature it has been custom to also use the matrix representation of this gauge field, denoted by $\tilde{A}_{\mu b}^a$. We will not use that gauge field in this paper. By the gauge field components we will always mean this kind of usual component field associated with associated Lie algebra generator $[T^a,T^b,\bullet]$, and will never use any matrix representation of this gauge field which is rather confusing as the indices are the same for these two enterily different objects.}
\bea
A_{\mu,ab}[T^a,T^b,\bullet] = A_{\mu,ab}[T^a,T^b] \Tr(\bullet) + 2A_{\mu,a\sharp}[T^a,\bullet] 
\eea
We may introduce new gauge fields as
\bea
A_a &=& 2A_{a\sharp}\cr
B_a T^a &=& A_{\mu,ab}[T^a,T^b]
\eea
Then
\bea
A_{\mu,ab}[T^a,T^b,\bullet] = B_{\mu,a} T^a \Tr(\bullet) + 2A_{\mu,a}[T^a,\bullet] 
\eea
We also introduce derivatives
\bea
D^A_{\mu} &=& \partial_{\mu} + A_{\mu}
\eea
With these preliminaries, we now find the supersymmetry variations of the membrane theory become
\bea
\delta \phi^I &=& i\bar{\epsilon} \Gamma^I \psi\cr
\delta \psi &=& \Gamma^{\mu}\Gamma_I\epsilon (D^{A}_{\mu}\phi^I + B_{\mu} \varphi^I) + \frac{1}{2}\Gamma_{IJK}\epsilon [\phi^I,\phi^J]\varphi^K\cr
\delta A_{\mu} &=& i\bar{\epsilon}\Gamma_{\mu}\Gamma_I \(\psi \varphi^I -\chi\phi^I\)
\eea
and 
\bea
\delta \varphi^I &=& i\bar{\epsilon} \Gamma^I \chi\cr
\delta \chi &=& \Gamma^{\mu}\Gamma_I \epsilon \partial_{\mu} \chi\cr
\delta B_{\mu} &=& i\bar{\epsilon} \Gamma_{\mu}\Gamma_I [\psi,\phi^I]
\eea

The restriction to Yang-Mills is now obtained by making the truncation
\bea
\chi &=& 0\cr
\varphi^I &=& R \delta^I_8\cr
B_{\mu} &=& 0
\eea
It may be noticed that this is consistent with the classical equations of motion provided
\bea
F_{\mu\nu} &=& \epsilon_{\mu\nu\rho} D^{\rho}\phi^8
\eea
This equation is now precisely what we need in order to descend to the supersymmetry variations of super Yang-Mills \cite{Bagger:2006sk}, 
\bea
\delta \phi^I &=& i\bar{\epsilon} \Gamma^I \psi\cr
\delta \psi &=& \frac{1}{2}\Gamma^{\mu\nu}\Gamma_{(10)}\epsilon F_{\mu\nu} + \Gamma^{\mu}\Gamma_M\epsilon D^{A}_{\mu}\phi^M + \frac{1}{2}\Gamma_{IJ}\Gamma_{(10)} \epsilon [\phi^M,\phi^N]\cr
\delta A_{\mu} &=& i\bar{\epsilon}\Gamma_{\mu}\Gamma_{(10)} \psi 
\eea
and we are back to the point from where the membrane theory originated. It arose from trying to up-lift precisely these supersymmetry variations to $SO(8)$ covariant variations.

\section{Other options?}\label{six}
Given that the membrane bracket is the one we have specified in Eq (\ref{bracket}), it is clear that the fields should take values in a centrally extended Lie algebra, since otherwise we do not get any non-vanishing traces. Could we get a non-trivial theory by taking a trivial central extension of the type $SU(N)\times U(1)$? If the central extension is trivial, then the $U(1)$ fields will obey free equations of motion (since the central element commutes with anything). In this sense the trivially centrally extendend membrane theory is trivial, and is probably just isomorphic to Yang-Mills theory upon a field redefinition. To obtain a non-trivial interacting theory we must have a non-trivial central extension (given the assumed form of the three-bracket), and this means that we must go to an infinite-dimensional gauge group since there are no non-trivial central extensions of finitie Lie algebras. This then is another motivation why loop space seems to arise for the $M2$ theories --  it seems to be the only way one can construct a non-trivial membrane theory.

It would certainly be nice if one could find a local field theory living on the membrane. Indeed there is one such local field theory, which corresponds to $SO(4)$ gauge group. This theory is constructed using a different three-product, given by
\bea
[a,b,c] = G_5 abc \pm antisym.
\eea
and the scalar fields and fermions are then given as
\bea
X^I &=& X^I_i G^i\cr
\psi &=& \psi_i G^i
\eea
Here $G^i$ are coordinates on a fuzzy three sphere, that generalizes the four-dimensional gamma matrices $\gamma^i$, and $G_5$ is then the analog of $\gamma_5$. 

It is not known if this theory can be reduced to Yang-Mills theory. It does not seem to be possible to generalize this $SO(4)$ theory to any higher (or lower) rank gauge groups. The algebraic structure (the Fundamental Identity together with antisymmetric structure constants of the three-algebra) of the membrane theory does not seem to admit any interesting generalizations, as we show in Appendix \ref{C}.

At last, let us show the connection between our $SU(2)\times SU(2)$ approach and the existing $SO(4)$ approach taken by Basu and Harvey in \cite{Basu-Harvey}. Our Nahm generators should be related to the (loop space generalization of the) Basu-Harvey generators $T^i$ as
\bea
T^I(s) &=& \lambda^I_{ij} [T^{i}(s),T^{j}(s)]\cr
t^I(s) &=& \lambda^I_{ij} [G^{i}(s),G^{j}(s)]\label{gf}
\eea
and analogously for the tilde generators. We find that
\bea
T^i(s) = \frac{G^i(s)}{\sqrt{z(s)\mp 1}}
\eea
satisfies the loop space generalization of the Basu-Harvey's equation
\bea
\frac{\delta T^i(s)}{\delta z(t)} \sim \epsilon_{ijkl} [T^j(s),T^k(t),T^l].
\eea
provided that $G^i(s)$ satisfies the loop space fuzzy sphere equation
\bea
\delta(s-t)G^i(s) \sim \epsilon_{ijkl}[G^j(s),G^k(t),G^l].\label{fg}
\eea
Here $T^i := \int ds T^i(s)$. It is a quite complicated story how the proportionality constant is determined, which depends on the choice of reducible representation of $SO(4)$ via the integer $n$. This can be found in \cite{Basu-Harvey}(v3). 

Moreover we get
\bea
T^I(s) = \frac{t^{I}(s)}{z(s)\mp 1}
\eea
which satisfies the loop space Nahm equation
\bea
\frac{\delta T^I(s)}{\delta z(t)} = \epsilon_{IJK} [T^I(s),T^J(t)].
\eea
since (by a slight extension of the short computation in footnote $1$ in the Introduction), 
\bea
4i\delta(s-t) t^I(s) = \epsilon_{IJK}[t^J(s),t^K(t)]
\eea
follows as a consequence of Eq's (\ref{fg}) and (\ref{gf}). 

Let us finally show the general implication of the Basu-Harvey equation (for the zero mode part for simplicity) here. That is, we show that 
\bea
\frac{dT^i}{dz} &=& \frac{1}{2(n+2)}\epsilon_{ijkl}G_5T^jT^kT^l
\eea
implies 
\bea
\frac{dT^I}{dz} &=& i\epsilon_{IJK}T^JT^K
\eea
where $T^I=\lambda^I_{ij}T^iT^j$. We compute the left-hand side,
\bea
\frac{dT^I}{dz} &=& \lambda^I_{ij}\(\frac{dT^i}{dz}T^j + T^i \frac{dT^j}{dz}\)\cr
&=& \frac{1}{2(n+2)}\(\lambda^I_{im}\epsilon_{ijkl} + \lambda^I_{ij}\epsilon_{imkl}\) G_5 T^jT^kT^lT^m\cr
&=& \frac{1}{2(n+2)}i\epsilon_{IJK}\lambda^J_{ij}\lambda^K_{kl} \(\P_+ - \P_-\) T^iT^jT^kT^l\cr
&=& i\epsilon_{IJK}\lambda^J_{ij}\lambda^K_{kl} T^iT^jT^kT^l\cr
&=& i\epsilon_{IJK} T^JT^K
\eea
In the second step we have assumed that $\{G_5,T^i\} = 0$. In the third step we use Eq (\ref{qwer}) to rewrite the 't Hooft matrices. In fourth step we used $G_5 = \P_+ - \P_-$ and $\P_+ \lambda^I_{ij}G^{ij} = (n+3) t^I$, $\P_- \lambda^I_{ij}G^{ij} = -(n+1)t^I$. The definition of $\P_{\pm}$ is found in \cite{Ramgoolam:2001zx} (where it is denoted as $\P_{{\cal{R}}_{\pm}}$, reflecting the fact that this is a projector on the representations ${{\cal{R}}}_{\pm}$ of $so(4)$).

The same computation can be done for the tilde variables. 

This shows that the Basu-Harvey equation implies the two Nahm equations. That means that the Basu-Harvey equation is identical to (or possibly a stronger equation than) two copies of the Nahm equation. 

We finally note that what was given the interpretation as a mysterious coupling constant $\lambda \sim \frac{1}{2(n+2)}$ in the Basu-Harvey equation in Ref  \cite{Basu-Harvey}, now simply disappears when we reformulate this equation as two Nahm equations, and hence this coupling constant need no explanation as it is no longer there.

\subsection*{Acknowledgements} I would like to thank Ulf Gran and Jakob Palmkvist for informing me about negative results in all their attempts to find explicit and physically relevant solutions to the Fundamental Identity Eq (\ref{fui}).

\newpage
\appendix
\section{The isomorphism $su(2)\oplus su(2)\simeq so(4)$}\label{A}
We first embed $SU(N)$ in $SO(2N)$ for any $N$. If $t^I$ denote generators of $SU(N)$, then the embedding in $SO(2N)$ is given as
\bea
t^I \mapsto \lambda^I := a^{\dag}_{i} (t^I)^{ij} a_{j}
\eea
where 
\bea
a_i &=& \frac{1}{2}\(\gamma^{2i-1} - i \gamma^{2i}\)
\eea
are annihilation operators constructed out of $SO(2N)$ gamma matrices, which thus satisfy the algebra
\bea
\{a_i^{\dag},a_j\} = \delta_{ij}
\eea
of $N$ creation and $N$ annihilation operators.

Now this gives an embedding of $SU(N)$ into the Dirac spinor representation of $SO(2N)$ because
\bea
a_i^{\dag} a_j = \frac{1}{2}\delta_{ij} - \frac{i}{2} \(M_{2i-1,2j-1} + M_{2i,2j}\) - \frac{1}{2} \(M_{2i-1,2j} - M_{2i,2j-1}\)
\eea
The term $\delta_{ij}$ does not contribute to the embedding matrices because the $SU(N)$ matrices are traceless. This result was now derived in the spinor representation where 
\bea
M_{ij} = \frac{i}{2}\gamma_{ij}.
\eea
But we may now take the $M_{jk}$ as generators of $SO(2N)$ in any representation. In particular we may take the defining vector representation.

Taking $N=2$, we obtain the 't Hooft matrices ($I=1,2,3$)
\bea
\lambda^I &=& \frac{1}{2}\epsilon^{IJK} M^{JK} - M^{I4}.
\eea
Of course $SO(4)\simeq SU(2)\times SU(2)$, and we can embed another $SU(2)$ as
\bea
\t \lambda^I &=& \frac{1}{2}\epsilon^{IJK} M^{JK} + M^{I4}
\eea
Conversely, the $\lambda^I$, $\t \lambda^I$ generate all of $SO(4)$. 

The 't Hooft matrices satisfy 
\bea
\lambda^I \lambda^J &=& i\epsilon^{IJK}\lambda^K + \delta^{IJ}\cr
\t \lambda^I \t \lambda^J &=& i\epsilon^{IJK} \t \lambda^K + \delta^{IJ}\cr
[\lambda^I, \t \lambda^J] &=& 0
\eea
and 
\bea
\lambda^I_{ij} \lambda^I_{kl} &=& -2\delta_{ij,kl} - \epsilon_{ijkl}\cr
\t \lambda^I_{ij} \t \lambda^I_{kl} &=& -2\delta_{ij,kl} + \epsilon_{ijkl}
\eea

From these relations we also derive the identities
\bea
\epsilon_{ijkm}\lambda^K_{ml} &=& i \epsilon_{IJK}\lambda^I_{ij}\lambda^J_{kl} - \delta_{ik}\lambda^K_{jl} + \delta_{jk}\lambda^K_{il} - \delta_{lk}\lambda^K_{ij} \cr
-\epsilon_{ijkm}\t \lambda^K_{ml} &=& i \epsilon_{IJK}\t \lambda^I_{ij}\t \lambda^J_{kl} - \delta_{ik}\t \lambda^K_{jl} + \delta_{jk}\t \lambda^K_{il} - \delta_{lk}\t \lambda^K_{ij}\label{qwer}
\eea
Also, we have
\bea
\Tr(\lambda^I\lambda^J) &=& 4\delta^{IJ}\cr
\Tr(\t \lambda^I\lambda^J) &=& 0.
\eea

\section{Verifying the ADHMN construction in loop space}\label{B}
Here we check the steps in the verification of the Nahm construction, following closely the steps in \cite{Weinberg:2006rq}. We make the ansatz
\bea
A_I(s) &=& -i\int [dz]  v^{\dag}_t \partial_I(s) v_t\cr
\phi(s) &=& \int [dz] z(s) v_t^{\dag} v_t 
\eea
We will suppress the multiplication sign over $t$. Here 
\bea
\Delta^{\dag}_s v_s &=& 0\cr
\int[dz] v_r^{\dag} v_r &=& 1
\eea
and 
\bea
\Delta_r(z) = \frac{\delta}{\delta z(r)} + \(X^I_r - T^I(z,r)\)\sigma^I,
\eea
\bea
\frac{\delta T^K(s)}{\delta z(r)} + \frac{i}{2}\epsilon_{IJK}[T^I(s), T^J(r)] &=& 0
\eea
which is the generalized Nahm equation. 

We verify that this solves the Bogomolnyi equation by first computing 
\bea
\frac{1}{2}\epsilon_{IJK}F_{IJ}(s,t) &=& -i\epsilon_{IJK}\(\int [dz] \partial_I(s)v^{\dag}_r \partial_J(t)v_r + \int [dz][dz']\underbrace{v_r^{\dag}\partial_{I}(s)v_r}_{-(\partial_I(s)v^{\dag}_r)v_r} v_{r'}^{\dag}\partial_{J}(t)v_{r'}\)\cr
&=& -i\epsilon_{IJK}\int [dz][dz'] \partial_{I}(s) v_r^{\dag}(z) F_{rr'}(z,z') \partial_J(t) v_{r'}(z')
\eea
Here
\bea
F_{rr'}(z,z') = \delta(z-z')\delta(r-r') - v_r(z)v^{\dag}_{r'}(z')
\eea
obeys
\bea
\int [dz'] \prod_{r'} F_{rr'}(z,z')F_{r'r''}(z',z'') = F_{rr''}(z,z'')
\eea
(Here $\delta(z)$ is defined as $\int [dz] \delta(z)\phi(z) = \phi(0)$, and $\delta(r)$ is defined with respect to multiplication, hence it should perhaps more appropriately be written as $e^{\delta(r)}$.) To see this, the normalization condition (\ref{norm}) is an essential ingredient. Hence $F$ is a projection operator onto the space orthogonal to the kernel of $\Delta^{\dag}$, and can be written as
\bea
F_{rr'}(z,z') = \Delta_r(z) G_{r,r'}(z,z') \Delta^{\dag}_{r'}(z')
\eea
where
\bea
G_{rs}(z,z') = (\delta(r-s)\delta(z-z')\Delta^{\dag}_r(z)\Delta_s(z'))^{-1}
\eea
We can then continue on our line and get
\bea
\frac{1}{2}\epsilon_{IJK}F_{IJ}(s,t) &=& -i\epsilon_{IJK} \int [dz][dz']\underbrace{\partial_I(s)v_r^{\dag}(z)\Delta_r(z)}_{\(\Delta^{\dag}_r(z)\partial_{I}(s)v_r(z)\)^{\dag}}G_{rr'}(z,z')\Delta^{\dag}_{r'}(z')\partial_J(t)v_{r'}(z')\cr
&=& -i\epsilon_{IJK} \int [dz][dz']v_r^{\dag}(z)\sigma_I \delta(s-r)  G_{rr'}(z,z') \sigma_J \delta(r'-t) v_{r'}(z')\nonumber
\eea
Finally we note that $G_{rr'}(z,z')$ is diagonal when the Nahm equation is obeyed, hence we can commute $\sigma_I$ with $G$ and use $[\sigma_I,\sigma_J] = 2i\epsilon_{IJK}\sigma_K$ to get
\bea
&=& 2 \int [dz][dz'] v_s^{\dag}(z) \sigma_K G_{st}(z,z') v_t(z')
\eea

Let us now turn to the right-hand side,
\bea
D_I(s) \phi(t) &=& \int [dz] z(t) D_I(s) (v^{\dag}_rv_r)\cr
&=& \int [dz][dz'] \partial_I(s) v^{\dag}_r(z) F_{rr'}(z,z') z'(t)v_{r'}(z')\cr
&&+ \int [dz][dz'] z(t) v_r^{\dag}(z) F_{rr'}(z,z') \partial_I(s) v_{r'}(z')
\eea
Then we again replace $F_{rr'}(z,z') = \Delta_r(z) G_{rs}(z,z') \Delta_r^{\dag}(z')$ and use 
\bea
\Delta^{\dag}_r(z) v_r(z) &=& 0\cr
\Delta^{\dag}_r(z) z(t) &=& \delta(t-r)\cr
\Delta^{\dag}_r(z) \partial_I(s) v_r(z) &=& -\delta(s-r) \sigma_I v_r(z)
\eea
and get immediately that
\bea
D_I(s) \phi(t) &=& 2\int [dz][dz'] v_s^{\dag}(z) \sigma_I G_{st}(z,z') v_t(z')
\eea
This concludes the verification that we have indeed constructed solutions to the Bogomolnyi equation
\bea
F_{IJ}(s,t) = \epsilon_{IJK}D_K(s)\phi(t).
\eea

\section{No-go argument}\label{C}
In this appendix we make it plausible that there are no non-trivial finite-dimensional extensions of the $SO(4)$ membrane theory within the class of semi-simple Lie algebras, like $SU(N)$ if we require a totally antisymmetric structure constants $f^{abcd}$ of the three-algebra, as defined below. 

\subsection{The associated Lie algebra}
Assume the $T^a$ generate a three-algebra
\bea
[T^a,T^b,T^c] = f^{abc}{}_d T^d
\eea
where the three-bracket is subject to the Fundamental Identity (FI)\cite{Bagger:2007jr}
\bea 
[ab[cde]] &=& [[abc]de] + [c[abd]e] + [cd[abe]]\label{fui}
\eea
By just assuming the antisymmetry property $[abc] = - [bac]$, we can write this in the following equivalent form
\bea
[ab[cde]] - [cd[abe]] &=& [[abc]de] - [[abd]ce]
\eea
which reflects the Lie algebra structure of the three-algebra. More clearly, if we define a linear operator as
\bea
t^{ab}(\bullet) = [T^a,T^b,\bullet] 
\eea
then the three-algebra says that
\bea
t^{ab}(T^c) = f^{abc}{}_d T^d
\eea
and then
\bea
[t^{ab},t^{cd}](T^e) &=& t^{ab}(f^{cde}{}_f T^f) - t^{cd}(f^{abe}{}_f T^f)\cr
&=& \(f^{abf}{}_g f^{cde}{}_f - f^{cdf}{}_g f^{abe}{}_f\) T^g
\eea
Applying the fundamental identity, we get
\bea
&=& \(f^{abc}{}_g f^{gde}{}_f - f^{abd}{}_g f^{gce}{}_f\)T^f\cr
&=& -2f^{ab[c}{}_g t^{d]g}(T^e)\label{C}
\eea
The commutator is again a linear operator, and multiplication being defined by composition of linear operators, is associative. Hence the commutator satisfies the Jacobi identity. 

Not all the $t^{ab}$ need to be linearly independent linear operators. Let us define
\bea
t^A := \Gamma^A_{ab} t^{ab}
\eea
in such a way that the $t^A$ become linearly independent and is the maximal such set. Then the commutator must close into this set by the above result. That is,
\bea
[t^A,t^B] = C^{AB}{}_C t^C
\eea

\subsection{More structure}
Let us define
\bea
t^{ab} := \Gamma^{ab}_A t^A
\eea
where $t^A$ is the maximal set of linearly independent such operators (the operators $[T^a,T^b,*]$ need not be linearly independent). That means that $\Gamma^{ab}_A$ viewed as a matrix with row-index $A$ and column index $ab$, need not be a square matrix. If so, then we can just find a one-sided inverse, 
\bea
\Gamma^{ab}_A \Gamma_{ab}^B = \delta_A^B
\eea
that is, if the index range $A$ is less than the index range $ab$. Using this, we may thus invert the above definition and get
\bea
t^A = \Gamma^A_{ab} t^{ab}
\eea
We now also define
\bea
t^{A,b}{}_{e} := \Gamma^A_{cd} f^{cdb}{}_e
\eea
and hence
\bea
t^A(T^b) = t^{A,b}{}_{e} T^e
\eea

Contracting the Fundamental Identity 
\bea
[t^{ef}(T^a),T^b,\bullet] - [t^{ef}(T^b),T^a,\bullet] = [t^{ef},t^{ab}](\bullet)
\eea
by $\Gamma^A_{ef}$, we get
\bea
[t^A(T^a),T^b,\bullet] - [t^A(T^b),T^a,\bullet] = [t^A,t^{ab}]
\eea
or
\bea
t^{Aa}{}_c t^{cb} - t^{Ab}{}_c t^{ca} = [t^A,t^{ab}]
\eea
or
\bea
\(t^{Aa}{}_c \Gamma^{cb}_C - t^{Ab}{}_c \Gamma^{ca}_C\)t^C = \Gamma^{ab}_C[t^A,t^C]
\eea
Now since $t^A$ are linearly independent, and generate a semisimple Lie algebra, they must be associated with a non-degenerated Killing form
\bea
\kappa^{AB} = \Tr(t^At^B)
\eea
which we can use to conclude from the above that
\bea
[t^A,\Gamma^B] = C^{AB}{}_C \Gamma^C
\eea

Conversely one may check that
\bea
f^{abc}{}_d = \Gamma^{ab}{}_A t^{Ac}{}_d.
\eea
satisfies the Fundamental Identity using the two equations
\bea
[t^A,t^B] &=& C^{AB}{}_C t^C\cr
[t^A,\Gamma^B] &=& C^{AB}{}_C \Gamma^C
\eea

This means that instead of trying to solve the Fundamental Identity, we may instead solve the two equivalent equations above. 

An obvious solution is to take 
\bea
\Gamma^{Aab} = t^{Aab}
\eea
In essence it is the only solution. To see this we make the ansatz
\bea
\Gamma_A = X_{AB} t^B
\eea
Then we find that
\bea
C^{AB'C}X^B{}_{B'} = C^{AB}{}_{C'}X^{C'}{}_C
\eea
Since we are interested in 
\bea
f^{abcd} = \Gamma^{Aab} t^{Acd} = X_{AB} t^{Aab} t^{Bcd}
\eea
being completely antisymmetric, we may assume that $X^{AB}=X^{BA}$. Then the condition we find is
\bea
[X,t^A] = 0
\eea
in the adjoint representation. Hence, by Schur's lemma, $X=1$ for a simple Lie algebra where the adjoint is an irrep. 

The adjoint representation is not irreducible for semisimple Lie algebras though, and in that case we can have $X=\lambda$ with different constants $\lambda$ in different irreps, and we now have that
\bea
f^{abcd} = X_{AB} (t^A)^{ab} (t^B)^{cd}.
\eea
This is now the most general solution of the Fundamental Identity. 

We now ask if this can become totally antisymmetric in $abcd$. It is manifestly so in $ab$ and $cd$ separately, as well as under exchange between the pairs. For complete antisymmetry in all indices, it thus suffices to find solutions such that
\bea
f^{abcd} + f^{acbd} = 0.
\eea

\subsection{$SU(2)\subset SO(4)$}
$SU(2)$ can be embedded into $SO(4)$ by taking
\bea
t^1 &=& M^{23} + M^{14}\cr
t^2 &=& M^{31} + M^{24}\cr
t^3 &=& M^{12} + M^{34}
\eea
where 
\bea
(M^{ab})_{cd} = \delta^{ab}_{cd}
\eea
are the standard generators of $SO(4)$ in an orthogonal basis, with respect to the invariant metric $\delta^{ab}$.

We then compute
\bea
\sum_{I=1}^3 (t^I)^{ab} (t^I)^{cd} &=& \delta^{23}_{ab}\delta^{23}_{cd} + \delta^{23}_{ab}\delta^{14}_{cd} + \delta^{14}_{ab}\delta^{23}_{cd} + \delta^{14}_{ab}\delta^{14}_{cd} + ...
\eea

For this particular case of $SO(4)$ there is room for another $SU(2)$,
\bea
\t t^1 &=& M^{23} - M^{14}\cr
\t t^2 &=& M^{31} - M^{24}\cr
\t t^3 &=& M^{12} - M^{34}
\eea
and we get cancelation of terms like $\delta^{ab}_{12}\delta^{cd}_{12}$ and get
\bea
(t^I)^{ab} (t^I)^{cd} - (\t t^I)^{ab} (\t t^I)^{cd} &=& \delta^{ab}_{23}\delta^{cd}_{14} + \delta^{ab}_{14}\delta^{cd}_{23}\cr
&+&\delta^{ab}_{31}\delta^{cd}_{24} + \delta^{ab}_{24}\delta^{cd}_{31}\cr
&+&\delta^{ab}_{12}\delta^{cd}_{34} + \delta^{ab}_{34}\delta^{cd}_{12}\cr
&=& \epsilon^{abcd}
\eea
that is, a completely antisymmetric tensor.

\subsection{$SU(N)\subset SO(M)$}
$SU(N)$ can be embedded into the vector representation of $SO(2N)$, but for $N>2$ we can not fit any more semi-simple group into $SO(2N)$. This means that we can not cancel the symmetric terms, like for instance $\delta^{ab}_{14}\delta^{cd}_{14}$. We may embed $SU(N) \times SU(N')\times ...$ in some $SO(M)$ where $M>2N$. But then there will always survive some symmetric terms that can not canceled. The analog of the 't Hooft matrices will sit as $2N \times 2N$ block matrices inside the $SO(M)$ $M\times M$ matrices (in the vector representation), which for $N>2$ will be strictly smaller than the $M\times M$ matrices. Therefore there will always be uncanceled symmetric terms that contributes to $f^{abcd}$ when $N>2$.

\subsection{Weaker conditions}
Could it be that one could relax the assumption that $f^{abcd}$ be totally antisymmetric and still get a sensible membrane theory? If we just assume the symmetries
\bea
f^{abcd} = f^{[ab][cd]} = f^{[cd][ab]}
\eea 
we could find many solutions to the Fundamental Identity. However if we assume a three-bracket
\bea
[a,b,c\} = -[b,a,c\}
\eea
with no other symmetries, then we must also assume that
\bea
[[a,b,c\},\bullet,d\} - [d,\bullet,[a,b,c\}\} \pm antisym = 0
\eea
in order to close SUSY up to gauge transformations (which must be defined using an antiosymmetrized three-bracket). But this condition leads to the condition that the antisymmetric part of the bracket $[\bullet,\bullet,\bullet\}$ satisfies the Fundamental Identity as well, and hence we are back at where we started, which is to find non-trivial solutions of the Fundamental Identity for the totally antisymmetric three-bracket. 

We may instead relax the assumption that 
\bea
(t^A)^{ab}
\eea
be antisymmetric. That is, we relax the assumptions that $f^{abcd}$ is totally antisymmetric, and instead just assume the it is antisymmetric in its three first indices, $f^{[abc]d}$. This will not affect the supersymmetry variations of the membrane since these are introduced without any reference to the trace form $\Tr(T^aT^b)$. Then the Fundamental Identity becomes equivalent with the equations 
\bea
[t^A,t^B] &=& C^{AB}{}_C t^C\cr
2(t^A\Gamma^B)^{[ab]} &=& C^{AB}{}_C \Gamma^{C,ab}
\eea
One infinite set of solutions to these equations is provided by letting $A=ef$ and then 
\bea
t^{ef,c}{}_d &=& \delta^{[e}_{\sharp}F^{fc]}{}_d\cr
\Gamma^{ab}_{ef} &=& \delta^{ab}_{cd}\label{awata}
\eea
where $F^{ab}{}_c$ are structure constants in any Lie algebra, thus satisfying the Jacobi identity.

The three-bracket now becomes
\bea
[a,b,c] = [a,b]\Tr(c) + [b,c]\Tr(a) + [c,a]\Tr(b).
\eea
This solution to the Fundamental Identity was found in \cite{Awata:1999dz}.

The Lie algebra generators are $t^{a\sharp}$ together with the generators $t^{ab}$. The algebra of these generators is
\bea
[t^{ab},t^{cd}] &=& 0\cr
[t^{ab},t^{c\sharp}] &=& F^{ab}{}_d t^{cd}\cr
[t^{a\sharp},t^{b\sharp}] &=& -F^{ab}{}_c t^{c\sharp}
\eea
In general, we (naively) read off the Lie algebra structure constants from Eq (\ref{C}) as
\bea
C^{ab,cd}{}_{ef} &=& 2 f^{ab[c}{}_{[e} \delta^{d]}_{f]}
\eea
It seems like we should antisymmetrize this in $ab,cd$. This is not manifestly antisymmetric because we need to contract by $t^{ef}$ and use FI to show this antisymmetry. So when we drop $t^{ef}$ we get structure constants that are not manifestly antisymmetric. Hopefully we can restore the Lie algebra structure constants just by antisymmetrizing in $ab,cd$, though we have not really checked the Jacobi identity gets satisfied by this prescription. Of course this can be verified explicitly case by case. For each solution we find to FI we can check that we get structure constants of a Lie algebra. So far we have found no counterexample, i.e. there appears to be no case where the FI does not imply a Lie algebra structure.

If we also assume the existence of a metric $\Tr(T^aT^b) = \delta^{ab}$ that can be used to raise the fourth index on $f^{abc}{}_d$, we can then also raise $ef$ by 
\bea
\kappa^{ef,gh} = f^{efgh}
\eea
 We then get
\bea
C^{ab,cd,ef} &=& -2f^{ab[c}{}_{g} f^{d]efg}.
\eea
This is manifestly antisymmetric under exchange of $ab$ and $ef$. As a consequence of earlier antisymmetrization prescription in $ab$ and $cd$, this now becomes a totally antisymmetric tensor in all three multi-indices. This fact reflects invariance propery of this metric, 
\bea
\kappa([t^{cc'},t^{aa'}]t^{bb'}) + \kappa(t^{aa'}[t^{cc'},t^{bb'}]) &=& 0.
\eea
We now notice that Lie algebra metric $\kappa^{AB} = f^{abcd}$ and structure constants $C^{ABC} = C^{ab,cd,ef}$ are precisely those tensors that figures in the Bagger-Lambert action. We thus see that it is possible to express their Chern-Simons action as a standard gauge theory, in terms of Lie algbra generators, once one accepts that the Lie algebra metric to be used is $\kappa^{ab,cd} = f^{abcd}$ rather than $\delta^{ab,cd}$. \footnote{This was a comment by Edward Witten.}

Now, as the only case where an action can be written down is for $SO(4)$, we can also rely on the relation $SO(4) = SU(2)\times SU(2)$ and write the Bagger-Lambert action as a usual gauge theory in terms of a two $SU(2)$ gauge fields as in \cite{VanRaamsdonk:2008ft}, \cite{Berman:2008be}.

No obvious action can be written for the theory associated with structure constants given by (\ref{awata}). One may write down an action for the theory in the limit that all three-brackets vanish, $[\bullet,\bullet,\bullet] = 0$ though. We can consider two different situations: either we put $F^{ab}{}_c = 0$, or we put the $U(1)$ fields to zero. If we put the $U(1)$ fields to zero then that theory is governed by the following supersymmetric Lagrangian
\bea
-\frac{1}{2}D^{\mu}\phi^{Ia}D_{\mu}\phi^I_a + \frac{i}{2} \bar{\psi}^a \Gamma^{\mu} D_{\mu} \psi_a + \frac{1}{\lambda}\epsilon^{\mu\nu\lambda} \delta^{ab} A_{\mu a}\partial_{\nu}A_{\lambda b}
\eea
Here $D_{\mu} X_a = \partial_{\mu} X_a + F^{bc}{}_{a} A_{\mu b} X_c$. Supersymmetry variation for the gauge field is $\delta A_{\mu} = 0$, and for the matter fields it is as usual. 

If we instead include the $U(1)$ fields $\varphi$ and $\chi$ and the associated gauge field $B_{\mu}$, then we will get a different but still non-trivial theory, even in the case when all three-brackets vanish, that is, even when $F^{ab}{}_c = 0$. For instance, we find that the fermionic equation of motion reads
\bea
\Gamma^{\mu}\(\partial_{\mu} \psi_a + B_{\mu a} \chi\) &=& 0.
\eea
and the action appears to be
\bea
&&-\frac{1}{2}(D^{\mu}\phi^{I})^a(D_{\mu}\phi^I)_a + \frac{i}{2} \bar{\psi}^a \Gamma^{\mu} (D_{\mu} \psi)_a + \frac{1}{\lambda}\epsilon^{\mu\nu\lambda} \delta^{ab} B_{\mu a}\partial_{\nu}B_{\lambda b}
\eea
where now the covariant derivative is given by
\bea
(D_{\mu} \phi)_a &=& \partial_{\mu} \phi_a + B_{\mu a}\varphi
\eea
This action thus already involves a subtle coupling of the $U(1)$ fields to the $SU(N)$ fields. It could be interesting to have at least one situation where such an action exists, to better understand what such a coupling between $U(1)$ fields and $SU(N)$ could mean physically. For instance one could integrate out the $U(1)$ fields to get an effective $SU(N)$ action. 

The interactions governed by $F^{ab}{}_c$ can then be incorporated in perturbation theory in the interaction picture as an interacting Hamiltonian.

\end{document}